\documentclass{article}
\usepackage{amsmath}

\input{tcilatex}

\begin{document}

\author{\textbf{Howard E. Brandt} \\
U.S. Army Research Laboratory, Adelphi, MD\\
hbrandt@arl.army.mil \and \textbf{John M. Myers} \\
Gordon McKay Laboratory, Harvard University, Cambridge, MA\\
myers@deas.harvard.edu}
\title{\textbf{Expanded Conclusive Eavesdropping in Quantum Key Distribution}%
}
\maketitle

\begin{abstract}
The paper [Howard E. Brandt, "Conclusive eavesdropping in quantum key
distribution," J. Opt. B: Quantum Semiclass. Opt. \textbf{7} (2005)] is
generalized to include the full range of error rates for the projectively
measured quantum cryptographic entangling probe, and also the full range of
inconclusive rates for the entangling probe measured with the POVM receiver.


\end{abstract}

\section{INTRODUCTION}

A design was recently given for an optimized entangling probe attacking the
BB84 (Bennett-Brassard 1984) protocol of quantum key distribution and
yielding maximum Renyi information to the probe \cite{PRA-05}, \cite%
{ModOp-05}. Probe photon polarization states become optimally entangled with
the signal states on their way between the legitimate transmitter and
receiver. Although standard von-Neumann projective measurements of the probe
yield maximum information on the pre-privacy amplified key, if instead the
probe measurements are performed with a certain positive operator valued
measure, then the measurement results are conclusive, at least some of the
time \cite{JOB-05}. It was assumed throughout that the error rate $E$
induced by the probe in the legitimate signal was such that $0\leq E\leq 1/4$
for the projectively measured probe. Here we extend the analysis to cover
the full range of theoretical interest, namely $0\leq E\leq 1/3.$

\section{GENERALIZED ENTANGLING PROBE}

In the present work a generalization is given to include the full range of
error rates, $0\leq E\leq 1/3$, for the projectively measured quantum
cryptographic entangling probe, and the full range of inconclusive rates, $%
0\leq R_{?}\leq 1$, for the entangling probe measured with the POVM
receiver. To accomplish this, the following sign choices must be made for
the probe parameter $\mu $ in Eqs. (26) and (27) of \cite{PRA-05}:%
\begin{equation}
\ \cos \mu =[(1+\eta )/2]^{1/2},\ \ \ \ 
\end{equation}%
\begin{equation}
\ \sin \mu =\text{sgn}(1-4E)[(1-\eta )/2]^{1/2},\ \ \ \ 
\end{equation}%
in which we define%
\begin{equation}
\ \text{sgn}(x)\equiv \left\{ 
\begin{array}{c}
1,\ \ \ x>0\ \ \ \ \  \\ 
\ \ 0,\ \ \ x=0\ \ \ \ \ \ \  \\ 
-1,\ \ \ x<0\ \ \ \ \ \ \ \ 
\end{array}%
\right. \ 
\end{equation}%
One also has the definition, Eq. (75) of \cite{PRA-05}:%
\begin{equation}
\eta \equiv \left[ 8E(1-2E)\right] ^{1/2}.
\end{equation}%
In this case, the probe states $\left\vert A_{1}\right\rangle $, $\left\vert
A_{2}\right\rangle $, $\left\vert \alpha _{+}\right\rangle $, $\left\vert
\alpha _{-}\right\rangle $, and $\left\vert \alpha \right\rangle $ of \cite%
{JOB-05} become:%
\begin{equation}
\left\vert A_{1}\right\rangle \equiv \left[ \frac{1}{2}(1+\eta )\right]
^{1/2}\left\vert w_{0}\right\rangle +\ \text{sgn}(1-4E)\left[ \frac{1}{2}%
(1-\eta )\right] ^{1/2}\left\vert w_{3}\right\rangle ,
\end{equation}%
\begin{equation}
\left\vert A_{2}\right\rangle \equiv \ \text{sgn}(1-4E)\left[ \frac{1}{2}%
(1-\eta )\right] ^{1/2}\left\vert w_{0}\right\rangle +\left[ \frac{1}{2}%
(1+\eta )\right] ^{1/2}\left\vert w_{3}\right\rangle ,
\end{equation}%
\begin{eqnarray}
\left\vert \alpha _{+}\right\rangle  &=&\left[ \left( 2^{1/2}+1\right)
\left( 1+\eta \right) ^{1/2}+\text{sgn}(1-4E)\left( 2^{1/2}-1\right) \left(
1-\eta \right) ^{1/2}\right] \left\vert w_{0}\right\rangle   \notag \\
&&+\left[ \text{sgn}(1-4E)\left( 2^{1/2}+1\right) \left( 1-\eta \right)
^{1/2}+\left( 2^{1/2}-1\right) \left( 1+\eta \right) ^{1/2}\right]
\left\vert w_{3}\right\rangle ,\ \ \ \ \ \ \ \ \ \ \ \ \ 
\end{eqnarray}%
\begin{eqnarray}
\left\vert \alpha _{-}\right\rangle  &=&\left[ \left( 2^{1/2}-1\right)
\left( 1+\eta \right) ^{1/2}+\text{sgn}(1-4E)\left( 2^{1/2}+1\right) \left(
1-\eta \right) ^{1/2}\right] \left\vert w_{0}\right\rangle   \notag \\
&&+\left[ \text{sgn}(1-4E)\left( 2^{1/2}-1\right) \left( 1-\eta \right)
^{1/2}+\left( 2^{1/2}+1\right) \left( 1+\eta \right) ^{1/2}\right]
\left\vert w_{3}\right\rangle ,\ \ \ \ \ \ \ \ \ \ \ \ \ 
\end{eqnarray}%
\begin{eqnarray}
\left\vert \alpha \right\rangle  &=&\left[ \text{sgn}(1-4E)\left( 1-\eta
\right) ^{1/2}-\left( 1+\eta \right) ^{1/2}\right] \left\vert
w_{0}\right\rangle   \notag \\
&&+\left[ \left( 1+\eta \right) ^{1/2}-\text{sgn}(1-4E)\left( 1-\eta \right)
^{1/2}\right] \left\vert w_{3}\right\rangle ,\ \ \ \ \ \ \ \ \ \ \ \ 
\end{eqnarray}%
\newline
respectively, where $\left\vert w_{0}\right\rangle $ and $\left\vert
w_{3}\right\rangle $ are the orthonormal basis states in the two-dimensional
Hilbert space of the probe. As in \cite{PRA-05}, the upper sign choice in
Eq. (23) of \cite{PRA-05} has been chosen. Note that Eqs. (5)-(9) are
consistent with Eqs. (5)-(7) and (11)-(14) of \cite{JOB-05} for $0\leq E\leq
1/4$, as must be the case.$\ \ $It then follows that Eqs. (1)-(4) of \cite%
{JOB-05}, along with Eqs. (7)-(9) above, now apply for $0\leq E\leq 1/3$.
(Note that $E=1/3\ $corresponds to complete information gain by the quantum
cryptographic entangling probe.) Also the probe and measurement
implementations remain the same (as in \cite{PRA-05}, \cite{ModOp-05}, \cite%
{JOB-05}) with the initial state of the probe now given by Eq. (6) or (21)
below. In obtaining the maximum Renyi information gain $I_{opt}^{R}$ by the
probe, Eq. (23) of \cite{JOB-05}, from Eqs. (7) and (8) above and Eqs. (23)
and (17) of \cite{HB-1} and the discussion following Eq. (75) of \cite%
{PRA-05}, one first has 
\begin{equation}
I_{opt}^{R}=\log _{2}(2-Q^{2}),
\end{equation}%
and one readily obtains for the overlap $Q$ of correlated probe states: 
\begin{equation}
Q=\frac{\left\langle \alpha _{+}|\alpha _{-}\right\rangle }{|\alpha
_{+}||\alpha _{-}|}=\frac{1+3\text{sgn}(1-4E)(1-\eta ^{2})^{1/2}}{3+\text{sgn%
}(1-4E)(1-\eta ^{2})^{1/2}}.
\end{equation}%
Then substituting Eq.\ (4) in Eq. (11), one obtains 
\begin{equation}
Q=\frac{1+3\text{sgn}(1-4E)((1-4E)^{2})^{1/2}}{3+\text{sgn}%
(1-4E)((1-4E)^{2})^{1/2}},
\end{equation}%
where we mean the positive square root; i.e. 
\begin{equation}
((1-4E)^{2})^{1/2}=|1-4E|.
\end{equation}%
On noting that 
\begin{equation}
\text{sgn}(1-4E)|1-4E|=1-4E,
\end{equation}%
and substituting Eqs. (13) and (14) in Eq. (12), one obtains 
\begin{equation}
Q=\frac{1-3E}{1-E}.
\end{equation}%
Finally, substituting Eq. (15) in Eq. (10), one obtains Eq. (23) of \cite%
{JOB-05}, namely, 
\begin{equation}
I_{opt}^{R}=\log _{2}\left[ 2-\left( \frac{1-3E}{1-E}\right) ^{2}\right] ,
\end{equation}%
for the full range of error rates, $0\leq E\leq 1/3$, as required.

From Eq. (31) of \cite{JOB-05}, one also concludes that the full range of
inconclusive rates $R_{?}$ apply, namely $0\leq R_{?}\leq 1$. In this
regard, it is useful, for the implementation of the entangling probe
incorporating the POVM receiver to measure the probe, to parameterize the
states $\left\vert A_{1}\right\rangle $ and $\left\vert A_{2}\right\rangle $%
, and $\eta $, and sgn$(1-4E)$ above, along with Eqs. (29) and (33) of \cite%
{JOB-05}, in terms of the inconclusive rate. Thus, using Eq. (31) of \cite%
{JOB-05}, it follows that the parameter $E$ is 
\begin{equation}
E=\frac{1-R_{?}}{3-R_{?}}.
\end{equation}%
Then the reflection coefficient $R_{1}$ in the POVM receiver, Eq. (29) of 
\cite{JOB-05}, becomes 
\begin{equation}
R_{1}=\frac{1-R_{?}}{1+R_{?}}.
\end{equation}%
Also, substituting Eq.\ (17) in Eqs. (3)-(6) above, one obtains 
\begin{equation}
\eta =\frac{2[2(1-R_{?}^{2})]^{1/2}}{3-R_{?}},
\end{equation}%
\begin{equation}
\left\vert A_{1}\right\rangle \equiv \left[ \frac{1}{2}(1+\eta )\right]
^{1/2}\left\vert w_{0}\right\rangle +\ \text{sgn}(3R_{?}-1)\left[ \frac{1}{2}%
(1-\eta )\right] ^{1/2}\left\vert w_{3}\right\rangle ,
\end{equation}%
\begin{equation}
\left\vert A_{2}\right\rangle \equiv \ \text{sgn}(3R_{?}-1)\left[ \frac{1}{2}%
(1-\eta )\right] ^{1/2}\left\vert w_{0}\right\rangle +\left[ \frac{1}{2}%
(1+\eta )\right] ^{1/2}\left\vert w_{3}\right\rangle .
\end{equation}%
Thus for the case of measurement of the probe with the POVM receiver,
according to Eq. (17), $E$ can be treated as a parameter ranging form 0 to
1/3, and determined by the set inconclusive rate $R_{?}$. Also, the
reflection coefficient $R_{1}$ of the POVM receiver must be set according to
Eq. (18), by the inconclusive rate. Finally, according to Eq. (21), the
initial state $\left\vert A_{2}\right\rangle $ of the probe can be tuned to
a set inconclusive rate of the POVM receiver.

Finally, it is important to emphasize that, if the photon loss rate, due to
attenuation in the key distribution channel between the probe and the
legitimate receiver, equals the inconclusive rate $R_{?}$, and only the
conclusive states are relayed by the probe to the legitimate receiver, then
the entangling probe together with the POVM receiver can obtain complete
information on the pre-privacy-amplified key, once the polarization bases
are announced in the public channel during reconciliation \cite{JOB-05}.
Also, to counter alteration in the attenuation due to the probe, the
legitimate channel may be replaced by a more transparent one. One may
therefore conclude that the BB84 protocol \cite{Bennett1} has a
vulnerability very similar to the well-known vulnerability of the B92
(Bennett 1992) protocol \cite{B92}, \cite{JOB-05}, \cite{RMP}. It is also
possible that the popular Ekert protocol \cite{Ekert} has a similar
vulnerability. It is important to emphasize that, because for the present
implementation one has $0\leq E\leq 1/3$, the inconclusive rate, according
to Eq. (17),\ can range here from 0 to 1, and can match a corresponding loss
rate in the channel connecting the probe to the legitimate receiver. If the
inconclusive rate $R_{?}$ is chosen to match the loss rate in the channel
connecting to the legitimate receiver, then the initial state of the probe
must be tuned (using a polarizer located between the single-photon source
and the target entrance port of the CNOT gate) to the value given by Eq.(21).

\section{CONCLUSION}

The conclusive entangling probe, which incorporates the POVM receiver to
measure the quantum cryptographic entangling probe, is generalized to
include a full range of inconclusive rates. It follows that the inconclusive
rate of the POVM receiver can match any photon loss rate in the key
distribution channel, and the standard BB84 and Ekert protocols of quantum
key distribution then have vulnerabilities analogous to the well-known
vulnerability of the standard B92 protocol.

\section{ACKNOWLEDGEMENTS}

This work was supported by the U.S. Army Research Laboratory and the Defense
Advanced Research Projects Agency.

\bigskip


\begin{thebibliography}{9}
\bibitem{PRA-05} H. E. Brandt, \textquotedblleft Quantum-cryptographic
entangling probe," Phys. Rev. A \textbf{71}, 042312(14) (2005).

\bibitem{ModOp-05} H. E. Brandt, "Design for a quantum cryptographic
entangling probe," to appear in J. Mod. Optics (2005).

\bibitem{JOB-05} H. E. Brandt, "Conclusive eavesdropping in quantum key
distribution," J. Opt. B: Quantum Semiclass. Opt. \textbf{7} (2005).

\bibitem{HB-1} H. E. Brandt, "Probe optimization in four-state protocol of
quantum cryptography," Phys. Rev. A \textbf{66}, 032303(16) (2002).

\bibitem{Bennett1} C. H. Bennett and G. Brassard, \textquotedblleft Quantum
cryptography: public key distribution and coin tossing,\textquotedblright\
in Proceedings of the IEEE International Conference on Computers, Systems,
and Signal Processing, Bangalore, India (IEEE, New York, 1984), pp. 175--179.

\bibitem{B92} C. H. Bennett, "Quantum cryptography using any two
nonorthogonal states," Phys. Rev. Lett. \textbf{68}, 3121-3124 (1992).

\bibitem{RMP} N. Gisin, G. Ribordy, W. Tittel, and H. Zbinden, "Quantum
cryptography," Rev. Mod. Phys. \textbf{74}, 145-195 (2002). (See p. 152.)

\bibitem{Ekert} A. Ekert, \textquotedblleft Quantum cryptography based on
Bell's theorem," Phys. Rev. Lett. \textbf{67}, 661-663 (1991).
\end{thebibliography}
\end{document}